\documentclass[prb,twocolumn,twoside]{revtex4}

\usepackage{dcolumn}
\usepackage{amsmath}

\usepackage{graphicx}
\usepackage{pslatex}

\def\eref#1{(\ref{#1})}

\def\etal{{\it{}et~al.}}
\def\av#1{\overline{#1}}
\def\bigstrut{\vphantom{\sum_{q_{q_q}}^d}}

\newlength{\figurewidth}
\begin{document}

\preprint{SFI working paper 01--11--066}

\title{Ego-centered networks and the ripple effect\\
\textit{--- or ---}\\
Why all your friends are weird}
\author{M. E. J. Newman}
\affiliation{Santa Fe Institute, 1399 Hyde Park Road, Santa Fe, NM 87501,
U.S.A.}
\date{November 5, 2001}

\begin{abstract}
  Recent work has demonstrated that many social networks, and indeed many
  networks of other types also, have broad distributions of vertex degree.
  Here we show that this has a substantial impact on the shape of
  ego-centered networks, i.e.,~sets of network vertices that are within a
  given distance of a specified central vertex, the ego.  This in turn
  affects concepts and methods based on ego-centered networks, such as
  snowball sampling and the ``ripple effect.''  In particular, we argue
  that one's acquaintances, one's immediate neighbors in the acquaintance
  network, are far from being a random sample of the population, and that
  this biases the numbers of neighbors two and more steps away.  We
  demonstrate this concept using data drawn from academic collaboration
  networks, for which, as we show, current simple theories for the typical
  size of ego-centered networks give numbers that differ greatly from those
  measured in reality.  We present an improved theoretical model which
  gives significantly better results.
\end{abstract}


\maketitle

\section{Introduction}
\label{intro}
\begin{figure}[b]
\resizebox{\figurewidth}{!}{\includegraphics{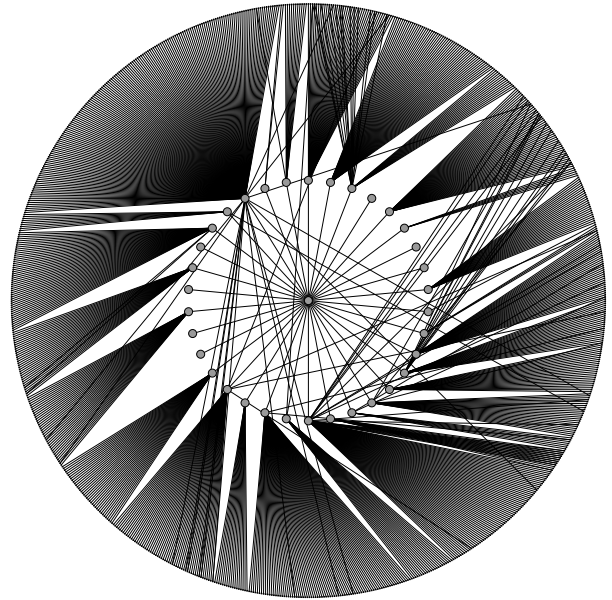}}
\caption{An ego-centered network of scientific collaborations, centered in
  this case on the author of this paper, represented by the vertex in the
  middle of the figure.  The two surrounding rings represent his
  collaborators, and their collaborators.  Collaborative ties between
  members of the same ring, of which there are many, have been omitted from
  the figure for clarity.}
\label{wheel}
\end{figure}

In social network parlance, an {\bf ego-centered network} (sometimes also
called a personal network) is a network centered on a specific individual
(generically ``actor''), whom we call the
ego.\cite{WF94,Scott00,Killworth90,Wellman93}  For example, Sigmund Freud
and all his friends would form an ego-centered network.  This network would
have radius~1, meaning we include everyone within distance~1 on the
friendship network of the central individual, Freud in this case.  If we
also included friends of friends in the network, it would have radius~2.
In Fig.~\ref{wheel} we show a radius-2 ego-centered network of scientific
collaborations.  The ego in this case is my own: the central vertex in the
figure represents me, the first ring of vertices around that my coauthors
on papers published within the last ten years, and the second ring their
coauthors.  As the figure shows, networks of this type can grow very
rapidly with radius.

Ego-centered networks are of interest for a number of reasons.  For
instance, in two recent papers, Bernard~\etal\cite{Bernard91,Bernard01}
address the following question.  Consider some subset of the population,
consisting of $e$ people.  They could be people in a particular demographic
or social group, or the people involved in a particular event.
How many of these $e$ people, if any, is the typical person likely to know?
As Bernard~\etal\ show, this is easy to calculate.  If the total population
who might be involved in the event is~$t$, then each member of that
population has a probability $p=e/t$ of being involved.  If the average
person knows $c$ other people, then the average number of those people who
were involved is simply $m=cp=ce/t$.  Bernard~\etal\ take the example of
the population of the United States, for which they estimate from previous
empirical studies that the average person has a social circle of about
$c=290$ people,\cite{Bernard91,Killworth90} and for which the total
population currently stands at around 280~million.  Thus the ratio
$t/c\simeq1\,000\,000$ in this case, giving the simple rule of thumb
\begin{equation}
m = {e\over1\,000\,000}.
\label{bernard}
\end{equation}
Simply stated, this equation says that the average individual living in the
United States is acquainted with about one person in a million out of the
country's total population.

As an example, let us apply the method to the problem of estimating how
many HIV positive individuals the average person in the US knows.  At the
time of writing, there were about $800\,000$ known cases of HIV in the US
(including those who have died).  The number of actual cases is probably
substantially greater than this and is estimated to be somewhere between
1.0 and 1.5~million.  To take a conservative figure, let us suppose that
the actual total is $e = 1$~million.  From Eq.~\eref{bernard}, we then
estimate that on average each member of the US population as a whole has or
had one acquaintance who is or was HIV positive.  It must be emphasized
that this is an average figure.  HIV positive individuals are not a uniform
sample of the population.  Nonetheless, Eq.~\eref{bernard} is expected to
give a correct population average of~$m$.

Now we want to extend this calculation one step further.  If a person has
no immediate friends in the group under consideration, how many of their
friends' friends are in this group?  Alternatively, one could rephrase the
question and enquire how many people in the population as a whole have one
or more friends of friends in the specified group: one can visualize a
group or event as the center of a set of ever widening circles of influence
in the social network.  Colloquially, this is what we call the ``ripple
effect.''  The two questions here are equivalent, but not identical.  In
this paper we speak in the language of the former, which focuses our
attention on the calculation of the number of actors two steps away from
the ego in an ego-centered network.

Unfortunately, the calculation of this number is not simple.  An
approximate solution is given in Ref.~\onlinecite{Bernard91} but, as we
will demonstrate, this solution misses some important features of real
social networks and as a result can give answers that are inaccurate.  The
crucial point is that in many networks there exist a small number of actors
with an anomalously large number of ties.  While it may appear safe to
ignore these actors because they form only a small fraction of the
population, we show that in fact this is not so.  Because of the way the
ripple effect works, this small minority has a disproportionately large
influence, and ignoring them can produce inaccurate estimates for the
figures of interest.  We show here how to perform calculations that take
these issues into account correctly.

The topic of this paper is also of interest in some other areas of social
network theory.  One such area is ``snowball sampling,'' an empirical
technique for sampling social networks that attempts to reconstruct the
ego-centered network around a given central actor.\cite{Erickson78,Frank79}
In this technique, the central actor is first polled to determine the
identities of other actors with whom he or she has ties.  Then those actors
are polled to determine their ties, and so forth, through a succession of
generations of the procedure.  The statistical properties of snowball
samples have been studied using Markov chain theory\cite{Heckathorn97} and
the technique has been shown to give good (or at least predictable) samples
of populations in the limit where a large number of generations of actors
is polled.  Unfortunately, in most practical studies only a small number of
generations is polled, and in this case, as we will see, the sample may be
biased in a severe fashion: snowball samples, like calculations of the
ripple effect, are highly sensitive to the presence in the population of a
small number of actors with an unusually large number of ties.


The outline of this paper is as follows.  In Section~\ref{2degrees} we
calculate exactly the expected number of network neighbors at distance two
from a central individual, in a network without transitive triples.  In
Section~\ref{transitivity} we show how the resulting expression is modified
when the network has transitivity, and in Section~\ref{examples} we apply
our theory to two example networks, showing that in practice it appears to
work extremely well.  In Section~\ref{concs} we give our conclusions.

\section{Friends of friends}
\label{2degrees}
So how do you estimate the number of people who are two steps away from you
in a social network (or indeed in a network of any kind)?
Bernard~\etal\cite{Bernard01} suggest the following simple method.  If each
actor in a network has ties to $c$ others on average, and each of those has
ties to $c$ others, then the average number of actors two steps away
is~$c^2$.  There are some problems with this however.  First, as pointed
out in Ref.~\onlinecite{Bernard01}, people who know one another tend to
have strongly overlapping circles of acquaintance, so that not all of the
$c$ people your friend knows are new to you---many of them are probably
friends of yours.  In other circles this effect is called network
transitivity\cite{WF94} or clustering,\cite{WS98} and it is also related to
the concept of network density.\cite{Granovetter76,WF94} Typically, the
mean number of people two steps away from an actor can be reduced by a
factor of two or so by transitivity effects.  Bernard~\etal\ allow for this
by including a ``lead-in factor''~$\lambda$ in their calculation.  We
discuss transitivity in more detail in Section~\ref{transitivity}.

Even if we ignore the effects of transitivity, however, there is a
substantial problem with the simple estimate of the number of one's second
neighbors in a social network.  By approximating this number as $c^2$ we
are assuming that the people we know are by and large average members of
the population, who themselves know average numbers of other people.  But
we would be quite wrong to make such an assumption.  The people we know are
anything but average.

Consider two (fictitious) individuals.  Individual~A is a hermit with a
lousy attitude and bad breath to the point where it interferes with
satellite broadcasts.  He has only 10 acquaintances.  Individual~B is
erudite, witty, charming, and a professional politician.  She has 1000
acquaintances.  Is the average person equally likely to know A and~B?
Absolutely not.  The average person is 100 times more likely to know B than
A, since B knows 100 times as many people.\footnote{Here and elsewhere in
this paper, we assume that ties between actors are symmetric.  In other
words, if A is friends with~B, then we assume also that B is friends
with~A.}  Extending this argument to one's whole circle of friends, it is
clear that the people one knows will, overall, tend to be people with more
than the average number of acquaintances.\footnote{The subtitle of this
paper would in fact be more accurate, though less catchy, if it read ``Why
most of your friends are popular.''}  This means that the total number of
{\em their\/} friends---the people two steps away---will be larger than our
simple estimate would suggest.  And as we will show, it may be very much
larger.

The fundamental concept that we need to capture here is that not all people
have the same number of acquaintances.  In the language of social network
analysis, there is a distribution of the degrees of vertices in the social
network.  (Recall that the degree of a vertex is the number of other
vertices to which it is directly connected.)  Let $k$ denote the degree of
a vertex and $p_k$ the degree distribution, i.e.,~the probability that a
vertex chosen uniformly at random from the network will have degree~$k$.
Thus, for example, the mean degree $c$ of a vertex is
\begin{equation}
c = \av{k} = \sum_{k=0}^\infty k p_k.
\label{avk}
\end{equation}
Degree distributions have been measured for a variety of networks, and in
many cases are found to show great
variation.\cite{AJB99,FFF99,ASBS00,Newman01a,Jeong01} It is certainly not
the case that vertices always have degree close to the mean (although they
may in some networks\cite{Marsden87}).  A clear example of this can be seen
in Fig.~\ref{wheel}, in which some vertices have degree only~1, while at
least one has degree greater than~100.

Now the fundamental point of this paper may be expressed as follows.  The
distribution of the degrees of the vertices to which a random vertex is
connected is {\em not\/} given by~$p_k$.  The probability that you know a
particular person is proportional to the number of people they know, and
hence the distribution of their degree is proportional to $kp_k$ and not
just~$p_k$.\cite{NSW01}  The correctly normalized distribution is thus
\begin{equation}
q_k = {kp_k\over\sum_{k=0}^\infty kp_k}.
\label{defsq}
\end{equation}

Now consider the number of vertices two steps away from a given vertex.
The probability $P(k_2|k_1)$ that this number is~$k_2$, given that the
number of vertices one step away is $k_1$, is
\begin{equation}
P(k_2|k_1) = \underbrace{\bigstrut
             \sum_{m_1=1}^\infty \sum_{m_2=1}^\infty \ldots\!
             \sum_{m_{k_1}=1}^\infty}_{\mbox{\small degrees of neighbors}}
             \hspace{2mm}\underbrace{\bigstrut\delta\!
             \left(\sum_{i=1}^{k_1}(m_i-1),k_2\right)}_{\mbox{\small
             degrees sum to $k_2$}} \>\prod_{j=1}^{k_1} q_{m_j},
\label{pk2k1}
\end{equation}
where $\delta(m,n)$ is~1 if $m=n$ and~0 otherwise.  Note the occurrence of
$m_i-1$ in this expression; the amount that your $i$th neighbor contributes
to the total number of your second neighbors is one less than his or her
degree, because one of his or her neighbors is you.  The overall
probability that the number of second neighbors is $k_2$ can then be
calculated by averaging Eq.~\eref{pk2k1} over~$k_1$:
\begin{equation}
P(k_2) = \sum_{k_1=0}^\infty p_{k_1} P(k_2|k_1).
\label{pk2}
\end{equation}
We want the mean value of $k_2$, which we will denote $c_2$, and this is
given by
\begin{equation}
c_2 = \av{k_2} = \sum_{k_2=0}^\infty k_2 P(k_2).
\label{avk2}
\end{equation}
Combining Eqs.~\eref{defsq}--\eref{avk2}, we thus arrive at the quantity we
are interested in:
\begin{eqnarray}
c_2 &=& \sum_{k_2=0}^\infty k_2 \sum_{k_1=0}^\infty p_{k_1}
        P(k_2|k_1)
     =  \sum_{k_1=0}^\infty p_{k_1} \sum_{k_2=0}^\infty k_2
        P(k_2|k_1)\nonumber\\
    &=& \sum_{k_1=0}^\infty p_{k_1}
        \sum_{m_1=1}^\infty \ldots\!\sum_{m_{k_1}=1}^\infty
        \sum_{k_2=0}^\infty k_2\>
        \delta\!\left(\sum_{i=1}^{k_1}(m_i-1),k_2\right)
        \prod_{j=1}^{k_1} q_{m_j}\nonumber\\
    &=& \sum_{k_1=0}^\infty p_{k_1} \sum_{i=1}^{k_1}
        \sum_{m_1=1}^\infty \ldots\!\sum_{m_{k_1}=1}^\infty
        (m_i-1) \prod_{j=1}^{k_1} q_{m_j}\nonumber\\
    &=& \sum_{k_1=0}^\infty k_1 p_{k_1}
        \left[\sum_{m=1}^\infty q_m\right]^{k_1-1}
        \sum_{k=0}^\infty (k-1) q_k\nonumber\\
    &=& \sum_{k=0}^\infty k(k-1)p_k
     =  \av{k^2} - \av{k}.
\label{bigeq}
\end{eqnarray}
This result
\begin{equation}
c_2=\av{k^2}-\av{k},
\label{mainres1}
\end{equation}
is the correction we were looking for to the simple estimate of the number
of vertices two steps away.\footnote{An alternative derivation of this
  result using probability generating functions can be constructed using
  results given in Ref.~\onlinecite{NSW01}.}  {\em The number of vertices
  two steps away is given by the mean square degree minus the mean degree.}
The important point to notice is that this expression depends on the
average of the square of a vertex's degree, rather than the square of the
average, as the simple estimate assumes.  If the degrees of vertices are
narrowly distributed about their mean, then these two quantities will be
approximately equal and the simple estimate will give roughly the right
result.  As mentioned above, however, many networks have broad degree
distributions, and in this case the average of the square and the square of
the average will take very different values.  In general, we can write
\begin{equation}
c_2 = \av{k^2} - \av{k} + (\av{k}^2 - \av{k}^2)
    = c^2 - c + \sigma^2,
\end{equation}
where $\sigma^2$ is the variance of the degree distribution and $c$ is, as
before, its mean.  Normally, $\sigma^2\gg c$ and so the difference between
the simple estimate $c^2$ and the true value of $c_2$ is about equal to the
variance.  In Section~\ref{examples} we give some examples of real networks
for which the variance is large---much larger than $c^2$ itself---and hence
for which the simple estimate gives poor results.

\section{Transitivity and mutuality}
\label{transitivity}
\begin{figure}
\resizebox{4.5cm}{!}{\includegraphics{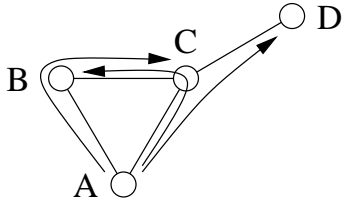}}
\caption{An illustration of the calculation of the clustering coefficient
  for a small network.  Vertex~A has three paths of length~2 leading from
  it, as marked.  Similarly vertices~B, C, and~D have~3, 2, and~2 such
  paths, for a total of $3+3+2+2=10$.  There is one triangle in the
  network.  Hence, from Eq.~\eref{defsc1}, the clustering coefficient is
  $6\times\frac{1}{10}=\frac35$.  Alternatively, one can count the number
  of connected triples of vertices, of which there are five, one each
  centered on vertices~A and~B, three on vertex~C, and none on vertex~D.
  Using Eq.~\eref{defsc2}, the clustering coefficient is then
  $3\times\frac15=\frac35$ again.}
\label{clustering}
\end{figure}

The calculation of the previous section is incomplete for a number of
reasons.  Chief among these is that it misses the effect of network
transitivity or clustering.  In most social networks, adjacent actors have
strongly overlapping sets of acquaintances.  To put this another way, there
is a strong probability that a friend of your friend is also your friend.
Transitivity can be measured by the quantity
\begin{equation}
C = {\mbox{$6\times$ number of triangles in the network}\over
     \mbox{number of paths of length two}}.
\label{defsc1}
\end{equation}
Here paths of length two are considered directed and start at a specified
vertex.  A ``triangle'' is any set of three vertices all of which are
connected to each of the others.  The factor of six in the numerator
accounts for the fact that each triangle contributes six paths of length
two to the network, two starting at each of its vertices.  This definition
is illustrated in Fig.~\ref{clustering}.  Simply put, $C$~is the
probability that a friend of one of your friends will also be your friend.

The quantity $C$ has been widely studied in the theoretical literature, and
its value has been measured for many different networks.  Watts and
Strogatz\cite{WS98} have dubbed it the {\bf clustering coefficient}.  It
is sometimes also known as the ``fraction of transitive triples'' in the
network.  Eq.~\eref{defsc1} is not in the form of the standard definition,
and so may not be immediately recognizable as the same quantity discussed
elsewhere.  The most commonly used definition is\cite{NSW01}
\begin{equation}
C = {\mbox{$3\times$ number of triangles in the network}\over
     \mbox{number of connected triples of vertices}},
\label{defsc2}
\end{equation}
where a ``connected triple'' means a vertex that is connected to an
(unordered) pair of other vertices.  It takes only a moment to convince
oneself that the two definitions are equivalent---see Fig.~\ref{clustering}
again.  (Note that paths are ordered in~\eref{defsc1} and triples are
unordered in~\eref{defsc2}, which accounts for an apparent difference of a
factor of two between the two definitions.)

What effect does clustering have on our calculation of the number $c_2$ of
second-nearest neighbors in the network?  Consider a vertex with degree $m$
lying in the first ``ring'' of our ego-centered network, i.e.,~one of the
immediate neighbors of the central vertex.  Previously we considered all
but one of this vertex's $m$ neighbors to be second neighbors of the
central vertex.  (The remaining one is the central vertex itself.)  This is
why the term $m-1$ appears in Eq.~\eref{bigeq}.  Now, however, we realize
that in fact an average fraction $C$ of those $m-1$ neighbors are
themselves neighbors of the central vertex and hence should not be counted
as second neighbors.  Thus $m-1$ in Eq.~\eref{bigeq} should be replaced
with $(1-C)\times(m-1)$.

Making this substitution in Eq.~\eref{bigeq} we immediately see that
\begin{equation}
c_2 = (1-C)(\av{k^2}-\av{k}).
\label{mainres2}
\end{equation}
This result is in general only approximate, because the probability of a
vertex having a tie to another in the first ring is presumably not
independent of the degrees $m_i$ of the other vertices.  As we show in the
following section however, Eq.~\eref{mainres2} gives considerably better
estimates of $c_2$ than our first attempt, Eq.~\eref{mainres1}.

But this is not all.  There is another effect we need to take into account
if we are to estimate $c_2$ correctly.  It is also possible that we are
over-counting the number of second neighbors of the central individual in
the network because some of them are friends of more than one friend.  In
other words, you may know two people who have another friend in common,
whom you personally don't know.  Such relationships create ``squares'' in
the network, rather than the triangles of the simple transitivity.  To
quantify the density of these squares, we define another
quantity\footnote{We choose the name ``mutuality'' because $M$ is also the
  average number of mutual friends of two actors who are distance~2 apart
  in the network.} which we call the {\bf mutuality}~$M$:
\begin{equation}
M = {\mbox{mean number of vertices two steps away}\over
     \mbox{mean paths of length two to those vertices}}.
\label{defsm}
\end{equation}
In words, $M$ measures the mean number of paths of length two leading to
your second neighbor.  Because of the squares in the network,
Eq.~\eref{mainres2} overestimates $c_2$ by exactly a factor of $1/M$, and
hence our theory can be fixed by replacing $m-1$ in Eq.~\eref{bigeq} by
$M(1-C)(m-1)$.

\begin{figure}
\resizebox{6cm}{!}{\includegraphics{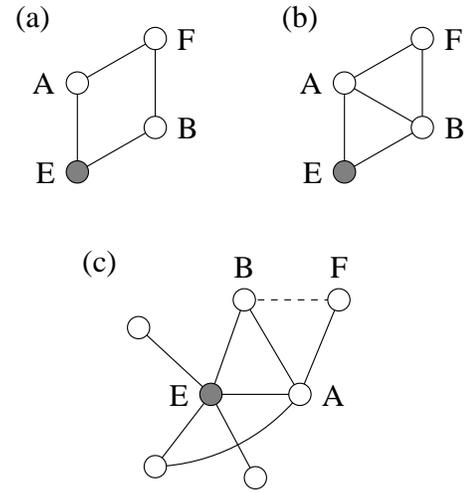}}
\caption{(a)~An example of an actor~(F) who is two steps away from the
  ego~(E, shaded), but is friends with two of E's friends (A~and~B).
  F~should only be counted once as a friend of a friend of~E, not twice.
  (b)~A similar situation in which A~and~B are also friends of one another.
  (c)~The probability of situation (b) can be calculated by considering
  this situation.  Since A is friends with both B and~F, the probability
  that B and F also know one another (dotted line), thereby completing the
  quadrilateral in~(b), is by definition equal to the clustering
  coefficient.}
\label{arrangement}
\end{figure}

But now we have a problem.  Calculating the mutuality $M$ using
Eq.~\eref{defsm} requires that we know the mean number of individuals two
steps away from the central individual.  But this is precisely the quantity
$c_2$ that our calculation is supposed to estimate in the first place.  Our
entire goal here is to estimate $c_2$ without having to measure it
directly, which would in any case be quite difficult for most networks.
There is however a solution to this problem.  Consider the two
configurations depicted in Fig.~\ref{arrangement}, parts~(a) and~(b).
In~(a), the ego, denoted E and shaded, has two friends A and~B, both of
whom know~F, although F is a stranger to~E.  The same is true in~(b), but
now A and~B are friends of one another also.  For many networks we find
that situation (a) is quite uncommon.  It is rare to find four people
arranged in a ring such that each knows two of the others, but none of the
four knows the person opposite them in the ring.  Situation (b) is much
more common.  And it turns out that we can estimate the frequency of
occurrence of (b) from a knowledge of the clustering coefficient.

Consider Fig.~\ref{arrangement}c.  The central actor~E has a tie with~A,
who has a tie with~F.  How many other paths of length two are there from~E
to~F?  Well, if E has $k_1$ neighbors, as before, then by the
definition~\eref{defsc2} of the clustering coefficient, A~will have ties to
$C(k_1-1)$ of them on average.  The tie between actors~A and~B in the
figure is an example of one such.  But now~A has ties to both B and~F, and
hence, using the definition of the clustering coefficient again, B~and~F
will themselves have a tie (dotted line) with probability~$C$.  Thus there
will on average be $C^2(k_1-1)$ other paths of length~2 to~F, or
$1+C^2(k_1-1)$ paths in total, counting the one that runs through~A.  This
is the average factor by which we will over-count the number of second
neighbors of~E because of the mutuality effect.  Substituting into
Eq.~\eref{bigeq}, we then conclude that our best estimate of $c_2$ is
\begin{equation}
c_2 = M(1-C)(\av{k^2}-\av{k}),
\label{mainres3}
\end{equation}
where the mutuality coefficient $M$ is given by
\begin{equation}
M = {\av{k/[1+C^2(k-1)]}\over\av{k}}.
\label{mutuality}
\end{equation}
Notice that both $1-C$ and $M$ tend to~1 as $C$ becomes small, so that
Eq.~\eref{mainres3} becomes equivalent to Eq.~\eref{mainres1} in a network
where there is no clustering, as we would expect.

In essence what Eq.~\eref{mutuality} does is estimate the value of $M$ in a
network in which triangles of ties are common, but squares that are not
composed of adjacent triangles are assumed to occur with frequency no
greater than one would expect in a purely random network.

To summarize, if we know the degree distribution and clustering coefficient
of a network---both of which can be estimated from knowledge of actors'
personal radius-1 networks---then we can estimate the number $c_2$ of
friends of friends the typical actor has using Eq.~\eref{mainres1},
\eref{mainres2}, or~\eref{mainres3}.  These three equations we expect to
give successively more accurate results for $c_2$.  Because we have
neglected configurations of the form shown in Fig.~\ref{arrangement}a and
because of approximations made in the derivation of Eqs.~\eref{mainres2}
and~\eref{mainres3}, we do not expect any of them to estimate $c_2$
perfectly.  As we will see in the following section however,
Eq.~\eref{mainres3} provides an excellent guide to the value of~$c_2$ in
practice, with only a small error (less than ten percent in the cases we
have examined).

\section{Example application}
\label{examples}
In this section we test our theory by applying it to two networks for which
can directly measure the mean number of second neighbors of a vertex and
compare it with the predictions of Eqs.~\eref{mainres1}, \eref{mainres2},
and~\eref{mainres3}.

Academic coauthorship networks are one of the best documented classes of
social networks.  In these networks the vertices represent the authors of
scholarly papers, and two vertices are connected by an edge if the two
individuals in question have coauthored a paper together.\footnote{In fact,
  coauthorship networks are most correctly represented as affiliation
  networks (i.e.,~bipartite graphs) of authors and papers, with edges
  connecting authors to the papers that they wrote or co-wrote.\cite{NSW01}
  Here, however, we deal only with the one-mode projection of the
  coauthorship network onto the authors.}  With the advent of comprehensive
electronic databases of published papers and preprints, large coauthorship
networks can be constructed with good reliability and a high degree of
automation.  Coauthorship networks are true social networks in the sense
that two individuals who have coauthored a paper are very likely to be
personally acquainted.  (There are exceptions, particularly in fields such
as high-energy physics, where author lists running to hundreds of names are
not uncommon.  We will not be dealing with such exceptions here, however.)

\begin{figure}[t]
\resizebox{\figurewidth}{!}{\includegraphics{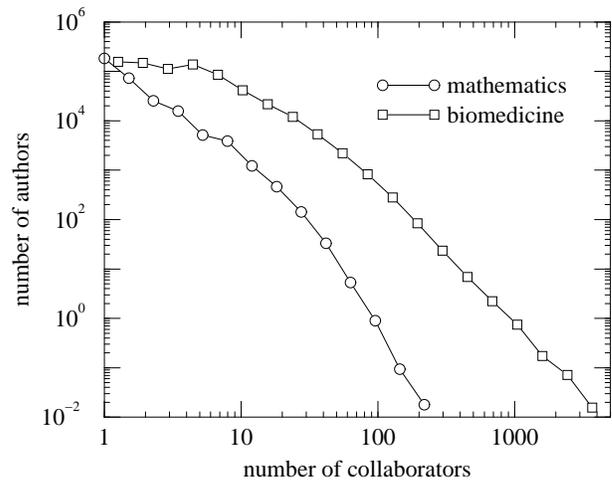}}
\caption{Degree distributions of the two academic coauthorship networks
  discussed in the text.  Axes are logarithmic.}
\label{degree}
\end{figure}

\begin{table*}
\setlength{\tabcolsep}{6pt}
\begin{tabular}{l|rrr|rrrr|r}
 & & & & \multicolumn{4}{c|}{estimate of $c_2$} \\
 network      & actors        & mean degree & clustering & simple & good
 & better & best   & actual $c_2$ \\
\hline
 mathematics  &    $253\,339$ &  3.92       & 0.150      & 15.4   & 47.9   & 40.7   & 33.8   & 36.4   \\
 biomedicine  & $1\,520\,251$ & 15.53       & 0.081      & 241.1  & 2006.0 & 1843.6 & 1357.6 & 1254.0 \\
\end{tabular}
\caption{Summary of results for collaboration networks of mathematicians
  and biomedical scientists.}
\label{summary}
\end{table*}

We examine two different coauthorship networks:
\begin{enumerate}
\item A network of collaborations between 1.5~million scientists in
  biomedicine, compiled by the present author\cite{Newman01a,Newman01b}
  from all publications appearing between 1995 and 1999 inclusive in the
  Medline bibliographic database, which is maintained by the National
  Institutes of Health.
\item A network of collaborations between a quarter of a million
  mathematicians, kindly provided to the author by Jerrold Grossman and
  Patrick Ion,\cite{GI95,DG99} who compiled it from data provided by the
  American Mathematical Society.
\end{enumerate}
In Fig.~\ref{degree} we show the degree distributions of these networks.
As the figure shows, neither is narrowly distributed about its mean.  Both
in fact are almost power-law in form, with long tails indicating that there
are a small number of individuals in the network with a very large number
of collaborators.  In the network of mathematicians, for instance, a
plurality (about a third) of individuals who have collaborated at all have
degree~1, i.e.,~have collaborated with only one other.  But there is one
individual in the network, the legendary Hungarian Paul
Erd\H{o}s,\cite{Hoffman98} who collaborated with a remarkable 502 others.
(This number is a lower bound; even though Erd\H{o}s died in 1996, new
collaborations of his are still coming to light through publications he
coauthored that are just now appearing in print.)

We have calculated the number of second neighbors of the average vertex in
these networks in five different ways: using the simple estimate discussed
in the introduction, using the three progressively more sophisticated
estimates, Eqs.~\eref{mainres1}, \eref{mainres2}, and~\eref{mainres3},
developed here, and directly by exhaustive measurement of the networks
themselves.  The results are summarized in Table~\ref{summary}.  As
expected, the simple method of estimating $c_2$, which assumes it to be
equal to the square of the mean degree, gives an underestimate for both
networks, by a factor of more than two for the mathematicians and more than
five for the biomedical scientists.  Moreover, we have been quite generous
to the simple method in this calculation, omitting from the formulas any
correction for transitivity, such as the lead-in factors discussed in the
introduction.  Including such a correction gives estimates of $c_2=10.9$
and $163.3$ for the two networks.  These estimates are too low by factors
of over 3 and 7 respectively---large enough errors to be problematic in
almost any application.

By contrast, the new method does much better.  The ``good'' and ``better''
estimates, Eqs.~\eref{mainres1} and~\eref{mainres2}, give figures of the
same general order of magnitude as the true result, and provide good
rule-of-thumb guides to the expected value of~$c_2$.  But the best
estimate, Eq.~\eref{mainres3}, making use of Eq.~\eref{mutuality} to
calculate the mutuality coefficient~$M$, does better still, giving figures
for $c_2$ which are within 8\% and 9\% of the known correct answers for the
mathematics and biomedicine networks respectively.  Clearly this is a big
improvement over the simple estimate.  Eq.~\eref{mainres3} appears to be
accurate enough to give very useful estimates of numbers of friends of
friends in real social networks.

\section{Conclusions}
\label{concs}
There are a number of morals to this story.  Perhaps the most important of
them is that your friends just aren't normal.  No one's friends are.  By
the very fact of being someone's friend, friends select themselves.
Friends are by definition friendly people, and your circle of friends will
be a biased sample of the population because of it.  This is a relevant
issue for many social network studies, but particularly for studies using
ego-centered techniques such as snowball sampling.

In this paper we have not only argued that your friends are unusual people,
we have also shown (in a rather limited sense) how to accommodate their
unusualness.  By careful consideration of biases in sampling and
correlation effects such as transitivity in the network, we can make
accurate estimates of how many people your friends will be friends with.
We have demonstrated that the resulting formulas work well for real social
networks, taking the example of two academic coauthorship networks, for
which the mean number of a person's second neighbors in the network can be
measured directly as well as estimated from our equations.

It is important to note however that application of the formulas we have
given requires the experimenter to measure certain additional parameters of
the network.  In particular, it is not enough to know only the mean number
of ties an actor has.  One needs to know also the distribution of that
number.  Measuring this distribution is not a trivial undertaking, although
some promising progress has been made
recently.\cite{Granovetter76,Killworth90,Bernard91} One must also find the
clustering coefficient of the network, which requires us to measure how
many pairs of friends of an individual are themselves friends.  This may
require the inclusion of additional questions in surveys as well as
additional analysis.

To return then to the question with which we opened this paper, can we
estimate how many friends of friends a person will have on average who fall
into a given group or who were involved in a given event?  If the number
involved in the event is $e$ as before, and the total population is~$t$,
then the number we want, call it $m_2$, is given by $m_2=c_2e/t$.  Thus,
once we have $c_2$ we can answer our question easily enough.  Using figures
appropriate for the United States and the simple estimate of $c_2$ that it
is equal to the square $c^2$ of the number of acquaintances the average
person has, we get $c_2=290^2=84100$, $t=280$~million, and $m_2=e/3330$.
As we have seen here, however, this probably underestimates the actual
figure considerably.  The real number could be a factor of five or more
greater than this formula suggests.  Unfortunately, as far as we know, the
necessary data have not been measured for typical personal acquaintance
networks to allow us to estimate $c_2$ by the methods described here.  In
particular, measurements of the clustering coefficient are at present
lacking.  We encourage those involved in empirical studies of these
networks to measure these things soon.

\begin{acknowledgments}
  The author thanks Michelle Girvan, Duncan Watts, and Elizabeth Wood for
  useful conversations, and Jerry Grossman, Oleg Khovayko, David Lipman,
  and Grigoriy Starchenko for graciously providing data used in the
  examples of Section~\ref{examples}.  This work was funded in part by the
  National Science Foundation under grant number DMS--0109086.
\end{acknowledgments}


\begin{thebibliography}{10}
\expandafter\ifx\csname url\endcsname\relax
  \def\url#1{\texttt{#1}}\fi
\expandafter\ifx\csname urlprefix\endcsname\relax\def\urlprefix{URL }\fi

\bibitem{WF94}
S.~Wasserman and K.~Faust, \textit{Social Network Analysis}. Cambridge
  University Press, Cambridge (1994).

\bibitem{Scott00}
J.~Scott, \textit{Social Network Analysis: A Handbook}. Sage Publications,
  London, 2nd edition (2000).

\bibitem{Killworth90}
P.~D. Killworth, E.~C. Johnsen, H.~R. Bernard, G.~A. Shelley, and C.~McCarty,
  Estimating the size of personal networks. \textit{Social Networks}
  \textbf{12}, 289--312 (1990).

\bibitem{Wellman93}
B.~Wellman, An egocentric network tale. \textit{Social Networks} \textbf{15},
  423--436 (1993).

\bibitem{Bernard91}
H.~R. Bernard, E.~C. Johnsen, P.~D. Killworth, and S.~Robinson, Estimating the
  size of an average personal network and of an event population: Some
  empirical results. \textit{Social Science Research} \textbf{20}, 109--121
  (1991).

\bibitem{Bernard01}
H.~R. Bernard, P.~D. Killworth, E.~C. Johnsen, G.~A. Shelley, and C.~McCarty,
  Estimating the ripple effect of a disaster. \textit{Connections}
  \textbf{24}(2), 18--22 (2001).

\bibitem{Erickson78}
B.~Erickson, Some problems of inference from chain data. In K.~F. Schuessler
  (ed.), \textit{Sociological Methodology 1979}, pp. 276--302, Jossey-Bass, San
  Francisco (1978).

\bibitem{Frank79}
O.~Frank, Estimation of population totals by use of snowball samples. In P.~W.
  Holland and S.~Leinhardt (eds.), \textit{Perspectives on Social Network
  Research}, pp. 319--348, Academic Press, New York (1979).

\bibitem{Heckathorn97}
D.~D. Heckathorn, Respondent-driven sampling: A new approach to the study of
  hidden populations. \textit{Social Problems} \textbf{44}, 174--199 (1997).

\bibitem{WS98}
D.~J. Watts and S.~H. Strogatz, Collective dynamics of `small-world' networks.
  \textit{Nature} \textbf{393}, 440--442 (1998).

\bibitem{Granovetter76}
M.~Granovetter, Network sampling: Some first steps. \textit{Am. J. Sociol.}
  \textbf{81}, 1287--1303 (1976).

\bibitem{AJB99}
R.~Albert, H.~Jeong, and A.-L. Barab\'asi, Diameter of the world-wide web.
  \textit{Nature} \textbf{401}, 130--131 (1999).

\bibitem{FFF99}
M.~Faloutsos, P.~Faloutsos, and C.~Faloutsos, On power-law relationships of the
  internet topology. \textit{Computer Communications Review} \textbf{29},
  251--262 (1999).

\bibitem{ASBS00}
L.~A.~N. Amaral, A.~Scala, M.~Barth\'el\'emy, and H.~E. Stanley, Classes of
  small-world networks. \textit{Proc. Natl. Acad. Sci. USA} \textbf{97},
  11149--11152 (2000).

\bibitem{Newman01a}
M.~E.~J. Newman, The structure of scientific collaboration networks.
  \textit{Proc. Natl. Acad. Sci. USA} \textbf{98}, 404--409 (2001).

\bibitem{Jeong01}
H.~Jeong, S.~Mason, A.-L. Barab\'asi, and Z.~N. Oltvai, Lethality and
  centrality in protein networks. \textit{Nature} \textbf{411}, 41--42 (2001).

\bibitem{Marsden87}
P.~V. Marsden, Core discussion networks of {A}mericans. \textit{American
  Sociological Review} \textbf{52}, 122--131 (1987).

\bibitem{NSW01}
M.~E.~J. Newman, S.~H. Strogatz, and D.~J. Watts, Random graphs with arbitrary
  degree distributions and their applications. \textit{Phys. Rev. E}
  \textbf{64}, 026118 (2001).

\bibitem{Newman01b}
M.~E.~J. Newman, Scientific collaboration networks: {I}. {N}etwork construction
  and fundamental results. \textit{Phys. Rev. E} \textbf{64}, 016131 (2001).

\bibitem{GI95}
J.~W. Grossman and P.~D.~F. Ion, On a portion of the well-known collaboration
  graph. \textit{Congressus Numerantium} \textbf{108}, 129--131 (1995).

\bibitem{DG99}
R.~de~Castro and J.~W. Grossman, Famous trails to {Paul Erd\H{o}s}.
  \textit{Mathematical Intelligencer} \textbf{21}, 51--63 (1999).

\bibitem{Hoffman98}
P.~Hoffman, \textit{The Man Who Loved Only Numbers}. Hyperion, New York (1998).

\end{thebibliography}
\end{document}